\documentstyle[11pt,aaspp4]{article}
\tighten
\eqsecnum

\begin{document}

\font\twelvei = cmmi10 scaled\magstep1 
       \font\teni = cmmi10 \font\seveni = cmmi7
\font\mbf = cmmib10 scaled\magstep1
       \font\mbfs = cmmib10 \font\mbfss = cmmib10 scaled 833
\font\msybf = cmbsy10 scaled\magstep1
       \font\msybfs = cmbsy10 \font\msybfss = cmbsy10 scaled 833
\textfont1 = \twelvei
       \scriptfont1 = \twelvei \scriptscriptfont1 = \teni
       \def\mit{\fam1 }
\textfont9 = \mbf
       \scriptfont9 = \mbfs \scriptscriptfont9 = \mbfss
       \def\bmit{\fam9 }
\textfont10 = \msybf
       \scriptfont10 = \msybfs \scriptscriptfont10 = \msybfss
       \def\bmsy{\fam10 }

\def\etal{{\it et al.~}}
\def\eg{{\it e.g.,~}}
\def\ie{{\it i.e.,~}}
\def\lsim{\raise0.3ex\hbox{$<$}\kern-0.75em{\lower0.65ex\hbox{$\sim$}}} 
\def\gsim{\raise0.3ex\hbox{$>$}\kern-0.75em{\lower0.65ex\hbox{$\sim$}}} 
\def\kms{~{\rm km~s^{-1}}}
\def\cm3{~{\rm cm^{-3}}}
\def\yr{~{\rm yr}}
\def\Msun{~{\rm M}_{\sun}}

\def\cf{{\it cf.~}}
\def\cc{CCs~}
\def\coll{ \tau_c}
\def\tma{ \tau_{ma}}
\def\tdr{ \tau_{dr}}
\def\tcr{ \tau_{cr}}
\def\tcs{ \tau_{cs}}
\def\ssp{c_{si}}
\def\roi{ \rho_i}
\def\ycm{ $Y_{cm}$~}
\def\pma{ \p_{max}}
\def\xcoor{{\it x}-coordinate~}
\def\ycoor{{\it y}-coordinate~}
\def\zcoor{{\it z}-coordinate~}
\def\ltsima{$\; \buildrel < \over \sim \;$}
\def\simlt{\lower.5ex\hbox{\ltsima}}
\def\gtsima{$\; \buildrel > \over \sim \;$}
\def\simgt{\lower.5ex\hbox{\gtsima}}

\title{On the Exchange of Kinetic and Magnetic Energy\\
   Between Clouds and the Interstellar Medium\altaffilmark{1}}

\author{Francesco Miniati,
        T. W. Jones}
\affil{School of Physics and Astronomy, University of Minnesota,
    Minneapolis, MN 55455}
\and     
\author{Dongsu Ryu}
\affil{Department of Astronomy \& Space Science, Chungnam National
    University, Korea}

\altaffiltext{1}{To appear in {\it The Astrophysical Journal} May 20, 1999 issue, 
Vol. 517 \#1}

\begin{abstract}

We investigate, through 2D MHD numerical simulations,
the interaction of a uniform magnetic field oblique to a moving interstellar
cloud. 
In particular we explore the transformation of cloud kinetic energy
into magnetic energy as a result of field line stretching. Some previous 
simulations 
have emphasized the possible dynamical importance of a ``magnetic shield''
formed around clouds when the magnetic field is perpendicular to the cloud 
motion (\cite{jrt96}, \cite{mrfj98}). It was not clear, however, how dependent 
those findings were to the assumed field configuration and cloud properties. 
To expand our understanding of this effect,
we examine several new cases by varing the magnetic
field orientation angle with respect to the cloud motion ($\theta$),
the cloud-background density contrast, and the cloud Mach number.

We show that in 2D and with $\theta$ large enough,
the magnetic field {\it tension} can become dominant in the dynamics of the 
motion of high density contrast, low Mach number clouds.
In such cases a significant fraction of cloud kinetic energy can be 
transformed into magnetic energy with the magnetic pressure at the cloud
nose exceeding the ram pressure of the impinging flow.
We derive a characteristic timescale, $\tma$, for this process of energy
``conversion''. 
We find also that unless the cloud motion is highly aligned to the
magneitc field, reconnection through
tearing mode instabilities in the cloud wake limit the formation of a strong
flux rope feature following the cloud. 
Finally we attempt to interpret some observational properties of the magnetic 
field in view of our results. 

\end{abstract}


\clearpage

\section{Introduction}

Magnetic fields represent an important component of the interstellar medium
(ISM)  and in many cases cannot be neglected when 
studying the processes taking place there. 
Numerous observations have been carried out and established 
important properties of the galactic magnetic field (\eg \cite{ve70},
\cite{he76}, \cite{zrs83}). 
Faraday rotation measures (\cite{gw63}, \cite{mb64}, \cite{vk75}) as well as
polarization measures (\cite{mf70}) have established the existence of
a large scale component of the galactic magnetic field, whereas
optical polarization of starlight and polarization of the background
synchrotron radio continuum have been used to determine its geometry 
(\cite{sp77}, \cite{gpb97}). 
Measurements of the Zeeman splitting of
the hyperfine 21 cm line of neutral hydrogen (\ion{H}{1}) have revealed 
a magnetic field strength of several $\mu$G for many interstellar
clouds (\cite{ve70}, 1989, \cite{he89}, \cite{mggh95}). 
In addition the magnetic field is often observed to be dynamically
relevant, with its energy being comparable to the cloud kinetic energy and, 
for self gravitating clouds, to their gravitational energy, too (\cite{he89},
\cite{ve89}, \cite{mggh95}). Since its importance was recognized
the magnetic field has been investigated in more and more detail.
The recent development of accurate magnetohydrodynamical (MHD) codes has provided 
an important new tool to compare and test analytical works. MHD numerical codes
also can investigate
beyond the necessary simplifications required to 
reduce an otherwise overwhelming mathematic complexity of
the problems under study. 

Magnetic fields may play an especially important role in the dynamics of cloud
motion through the ISM. Mac Low \etal (1994), for example, carried out 2D
numerical simulations of the interaction between a cloud a few times denser than 
the background medium and a plane shock wave and showed that even a modest 
magnetic field aligned with the direction of
the shock motion can help to stabilize the cloud against disruptive instabilities. This
results from stretching of the magnetic field into a wake that forms behind such a
cloud from its motion through the postshock plasma. That significantly reduces the
vorticity shed by the cloud, which helps to stabilize it. Jones \etal (1996)
demonstrated the same effect for clouds set into supersonic motion through the
ISM. The latter paper also examined in detail the magnetic field behavior during
supersonic cloud motion perpendicular to the prevailing magnetic field. In that
instance Jones \etal (1996) found that the magnetic field plays an even more dramatic
role, because field lines caught on the ``nose'' of the cloud become highly stretched
and form a ``magnetic shield'' with a magnetic pressure at least comparable to the gas ram
pressure. That also effectively quenches disruptive Rayleigh-Taylor and Kelvin-
Helmholtz instabilities (KHIs), so that the clouds maintain their full integrity. 
In addition,
this enhanced magnetic pressure at the front of the cloud hastens the cloud's 
deceleration
with respect to the ISM. They found this result even when the initial magnetic 
field
is quite modest (\ie wherever $\beta = p_g/p_b < 100$, for a Mach 10 cloud).
  
Most recently, Miniati \etal (1997, 1998)  have used 2D simulations to study 
head-on
collisions between mildly supersonic clouds in the ISM. Miniati \etal (1998) 
found a
dramatic influence of the magnetic field when it is perpendicular to the motion 
of
the clouds. If the magnetic shield seen by Jones \etal (1996) forms before the
collision takes place, it is seen to act as an almost elastic bumper between 
the two clouds,
preventing their direct contact and also their disruption during the 
interaction.

Those results clearly indicate a need to understand more completely 
the interactions between cloud motion and magnetic field. 
There are several simplifications in the previously
mentioned studies that need to be relaxed along the way to that understanding. 
The
two most obvious are geometrical. All of the above studies were 2D in the sense
that they assumed symmetry in one direction perpendicular to the cloud motion.
The second is the assumption that the magnetic field was either exactly aligned
with cloud motion or perpendicular to it. 

We are tackling the fully 3D study of
this problem in a different paper in preparation
(Gregori \etal 1998). In the transition from
2D to 3D, several new physical aspects appear, 
adding enormous complexity to the investigation process.
More realistic 3D MHD numerical simulations, although nowadays accessible
through large computers,
still remain much more expensive than 2D ones. Extending 
a computational box of size $N_x \times N_y$ to three dimensions 
with size $N_x \times N_y \times N_z$ implies an increase in the computational
cost (execution time) of the calculation by a factor about 
$\slantfrac{3}{2} ~N_z$. Since generally $N_{x,y,z}$ should be at least several
hundred to properly capture the dynamics, the tradeoff usually needed to carry
out 3D simulations is a reduction of either the computational box size or 
the resolution.
There is a lower limit on the box size set by the influence of the boundary
conditions. The latter are particularly important in this case because,
as we shall see, the evolution of the magnetic energy is significantly altered
by the outflow of regions of enhanced magnetic field. 
If, on the other hand, the resolution is reduced by a factor 
$r > 1$, then the cost of a 3D simulation with
respect to a 2D one is, instead, $\slantfrac{3}{2}~ N_z/r^4$. For example a 3D
calculation carried out on a $256^3$ grid ``costs'' 48 times the same 2D
calculation on a $512^2$ grid, compared to an increased cost factor of 768 on a
$512^3$ grid. Memory and data storage factors are comparable. Even the smaller
factor represents a major increased investment. In addition visualization and
3D analysis of data are much more difficult. 
For these reasons 2D calculations 
are still very worthwhile, especially when an exploratory parametrical study is to 
be carried out. This is the case in the present
circumstance, where we are investigating the role, in the 
cloud-magnetic field interaction, of several factors such as 
the Mach number
($M$), the angle between the cloud motion and the magnetic field lines
($\theta$) and the cloud-intercloud density contrast ($\chi$). 

Since the magnetic bumper seen in previous studies involves perpendicular
fields with clearly dominant features, the objective of this paper is to  
understand how nearly aligned the magnetic field can be with the cloud motion
and still form an effective magnetic bumper. That leads us naturally also to 
explore more generally the exchange between kinetic and magnetic energies during 
the motion of a 2D cloud through a conducting background. It is important for
this particular purpose to question the validity of the 2D approximation. 

First, in a 2D cartesian geometry we are limited to represent cylindrical clouds with
third axes perpendicular to the computational plane. Since
ISM clouds are observed to take many forms including filaments and sheets,
in this respect it is still appropriate to investigate the simple 2D 
case of a transverse cylinder. However, in 2D, instabilities along the symmetry
axis are inhibited. This means that our calculations are reliable only up to the
characteristic time for those instabilities to occur on a spatial scale
comparable to the typical cloud size. Since the most relevant instabilities for
dense clouds in the ISM develop on a timescale 
$\tcr =2~\chi^{1/2}/M~\tcs$ ($\tcs$ is the sound crossing time of the cloud,
see \S \ref{desp} for details), we will restrict the investigation of the 
evolution of the various cases to $t \lesssim$ several $\tcr$.

In addition it is
often remarked that in a realistic 3D situation the magnetic field 
lines would be free to slip by the cloud sides, therefore escaping from the
line stretching process. 
What happens in reality is quite complex and involves not
only the flow of the magnetized gas around the cloud, but also the backreaction
of the cloud to that flow and the embedded field. 
In Miniati \etal (1998) we addressed this issue briefly, 
pointing out the importance of the deformations undergone by the cloud during
its motion through a magnetized gas, and their role in the development
of the magnetic bumper. This issue will be investigated more thoroughly
in our forthcoming paper (Gregori \etal 1998). There we shall also address some
specific dynamical and morphological properties related to the
development of 3D instabilities. In particular, 3D Rayleigh-Taylor 
instabilities (RTIs) and the consequent
escape of magnetic flux tubes from the stretching region. In fact, the full
development of RTI in the direction perpendicular to both the magnetic field and
the cloud motion, suppressed in our 2D simulations, is able to break up the 
cloud after several $\tcr$, thus freeing the magnetic flux tubes 
from the stretching mechanism. This process was discussed in the context 
of Planetary Nebulae by Soker and Dgani (1997),
Dgani and Soker (1998) and Dgani (1998). 
In the meantime, in \S \ref{dis} we will
present some preliminary results about the growth of the magnetic energy in an
analogous 3D simulation of a cylinder moving through a transverse magnetic
field. In support of the present investigation we mention that those 3D 
results are in qualitative agreement with our 2D numerical study.

The paper outline is as follows:
\S \ref{desp} provides a brief background review relevant to this paper.
In \S \ref{code} the numerical setup is described. In \S \ref{res}
we present the results of our calculations and comment on their 
physical and astrophysical implications. In \S \ref{motiv} we attempt
to interpret some recent observational results based on our findings,
which are summarized in \S \ref{cs}.

\section{Hydrodynamics and Magnetohydrodynamics of Single Cloud Motion}
\label{desp}
A cloud set into motion within a lower density medium produces new 
structures in both the cloud and the background medium. 
Several features develop, each one characterized by a 
distinct timescale. In this section we will briefly review those features
most relevant to the current study, in order to facilitate the understanding
of the issues raised in this paper (for further details see \eg
\cite{mu93}; Jones \etal 1994, 1996; \cite{sck95}; \cite{mbr96}; \cite{vfm97}).  
The speed of the cloud 
($v_c$) is usually expressed in terms of the Mach number $M=v_c/c_{si}$, where 
$c_{si}$ is the sound speed in the unperturbed background medium. 
For standard \ion{H}{1} clouds, usually $M\ge 1$ (Spitzer 1978); that is,
the motion is supersonic. In purely gasdynamic cases, 
a bow shock forms in front of the cloud and reaches a constant standoff 
distance from it on the timescale $\tau_{bs}\sim 2R_c/v_c\equiv 
2\coll$ where $R_c$ is the cloud radius and $\coll$ is the cloud crossing time. 
This increases the gas density on 
the nose of the cloud and leads to a pressure there comparable to the ram pressure 
$\rho_i v_c^2$. In response to this, a ``crushing'' shock 
forms and propagates through the cloud with relative speed $v_{cs}\simeq 
v_{c}/\chi^{1/2}$,
where the density contrast parameter, $\chi=\rho_c/\rho_i$,
is the ratio of the cloud and intercloud medium densities. 
The timescale over which this shock crosses the cloud is 
\begin{equation}
\tcr = \frac{2R}{v_{cs}}=\frac{2R\chi^{\frac{1}{2}}}{v_c}
=2~\frac{\chi^{\frac{1}{2}}}{M}~\tcs
\end{equation}
where $\tcs=R_c/c_{si}$ is the cloud sound crossing time.
After a time $\sim \tcr$ the cloud structure has been significantly modified 
and is far from uniform. 
Indeed, as the cloud plows through the background gas, KHIs and RTIs develop on its 
boundary layer. The slowest growing but most effective modes in disrupting 
the cloud
are those characterized by the largest wavelength $\lambda\sim R$, and for 
$\chi\gg1$ their typical growth time is roughly the same as $\tcr$. 
Finally, as the ram pressure carries out work on the cloud, the cloud
velocity relative to the background medium is reduced. This takes place on the
timescale roughly given by (\cite{kmc94}) 
\begin{equation}
\tau_{dr} = \frac{3\chi R}{4v_{c}}\sim\frac{R\chi}{M c_{si}}
=\frac{\chi}{M}~\tcs.
\label{eqdr}
\end{equation}

The propagation of a cloud through a magnetized medium adds new parameters
to the characterization of the cloud evolution.
One is the strength of the magnetic field, expressed in terms of
\begin{equation}
\label{beta}
\beta = \frac{p_g}{p_B} = \frac{2}{\gamma}\left(\frac{M_A}{M}\right)^2
\end{equation}
where $p_g$ and $p_B=B^2/8\pi$ are the gas and magnetic pressure respectively
and $M_A=v_c/(B/\sqrt{4\pi\rho})$ is the Alfv\'enic Mach number.
The other is the field orientation, which in 2D calculations with $B_z=0$
is determined by the angle $\theta$ 
between the initial cloud velocity and the magnetic field lines.

For a large $\beta$, the beginning of the MHD cloud evolution
is similar to the HD case.
Over time, new features develop in response to the magnetic field. 
As mentioned earlier, 
numerical investigations have shown significant differences
between the cases of cloud motion parallel ({\it aligned})
or perpendicular ({\it transverse}) 
to the initial uniform magnetic field (\cite{jrt96},
\cite{mac94}). MacLow \etal (1994) 
studied the MHD interaction of a high Mach number ($M$=10) planar shock 
wave with a non radiative low density contrast ($\chi=10$) spherical
interstellar cloud. They investigated 
both aligned and, for one cylindrical cloud case, perpendicular
field geometries, although they mostly concentrate on the former geometry. 
On the other hand, Jones \etal (1996)
studied the propagation of an individual supersonic cylindrical
cloud characterized by the same 
parameters ($M=10,~\chi = 10$), through a magnetized medium. In both sets
of computations the initial magnetic field was uniform throughout the grid.
The studies found in common agreement  
that, in the aligned case, the field lines threading the cloud are 
pulled, stretched and folded around the cloud forefront, eventually 
reconnecting and generating new and more stable flux tubes in this
region.
Here the magnetic field never becomes dynamically dominant, although
it plays a relevant role in stabilizing the flow around the cloud. 
On the other hand, low pressure in the cloud wake and stretching of 
the field lines by the wake flow cause the magnetic field to prevail 
($\beta\le 10$) in this region (``flux rope'') behind the cloud 
(\cite{mac94}, \cite{jrt96}). In the shocked cloud problem
(\cite{mac94}) the magnetic field in the flux rope is amplified up to the 
postshock thermal pressure, whereas in the isolated cloud 
calculations (\cite{jrt96}) such amplification is smaller. 

When the cloud motion is transverse to the magnetic field,
field lines start to drape around the cloud surface, 
being compressed and, more importantly, stretched around the cloud nose.
In particular, because of stretching of the field lines, the magnetic
field in this region becomes eventually very strong, generating a 
stabilizing {\it magnetic shield} region
of small values of $\beta$ (\cite{jrt96}). 
The presence of the magnetic shield also
significantly alters the outcome of cloud collisions in 
the perpendicular field case (\cite{mrfj98}). Indeed, if the magnetic 
shield is strong enough, it acts like an elastic ``bumper'' 
reversing the clouds motion before they contact. 

Both of the studies involving perpendicular fields 
(\cite{jrt96}, \cite{mac94}) found
that the amplified field on the cloud nose becomes comparable to 
the ram pressure of the impinging flow, which is of the same order as
the thermal pressure of the postshock gas. We shall show below that 
those conclusions were implied by their choice of the cloud parameter
($\chi=10$ and $M=10$). Then the question raised is how one may more generally
relate the magnetic field properties to the initial conditions. 
Such question is quite complex and will be addressed in some more detail in \S
\ref{phi}.
We shall emphasize though that the enhancement of the magnetic energy is in 
general due primarily to the work done on field lines through stretching
by the moving cloud. Therefore, in principle, {\it the cloud initial kinetic 
energy, $\slantfrac{1}{2}~M_c v_c^2$, is the source and the upper limit 
for the magnetic energy}. 
Nevertheless, as we will see in the following, 
a cloud is not always in the right setting for its kinetic energy to be 
transformed into magnetic form.
However, when such conditions exist, the field enhancement maybe larger
than previously estimated by 
a factor in principle up to $\lesssim\chi$ which, for high density clouds, 
makes a significant difference. 

\section{Numerical Setup}
\label{code}

To explore these issues we have carried out numerical calculations with an 
ideal MHD code, based on the conservative, explicit, second-order TVD method
described in detail in Ryu \& Jones (1995), Ryu, Jones, \& Frank (1995)
and Ryu \etal (1998). 
Recent measurements have shown that \ion{H}{1} diffuse 
clouds are often characterized by 
high values of the electron fraction $x_e\simeq 10^{-2.7}-10^{-4.9}$
(\cite{mk95}).
Therefore, both the kinetic Reynolds
number, Re (=$vr/\nu$), and the magnetic Reynolds number, Re$_M$ 
(=$v\ell/\nu_M$), are expected to be large for velocities $v\sim$ a 
few km sec$^{-1}$ and spatial scales $\ell\sim 1$ pc (\cite{mk95}). 
The resulting ambipolar diffusion timescale is given by
\begin{equation}
\tau_{AD}=\left(\frac{L}{v}\right) Re_M=2.2\times 10^{9}
\left(\frac{n}{{10~\rm cm}^{-3}}\right)^2 \left(\frac{L}{{\rm pc}}\right)^2
\left(\frac{B}{\mu{\rm G}}\right)^{-2} \left(\frac{x_e}{10^{-3}}\right)
~{\rm yr};
\end{equation} 
meaning that the magnetic field should not decay through ion-neutral drift
during the cloud evolutionary time considered here. 
Ambipolar diffusion can also affect the shock wave
structure (Draine \& McKee 1993, and references therein). This is particularly
important inside neutral clouds where the ionization fraction $x_e$ can be as  
small as $\sim 10^{-5}$. As a result the ``shock transition'', across which both 
ions and neutrals undergo the ordinary ``jump conditions'', can become so long
that radiative cooling is relevant there (\eg Draine \& McKee 1993). 
However, this process only affects the thickness 
and structure of the shock wave, but not the overall jump conditions, which 
are set by conservation equations. These jump conditions are always correctly 
calculated by any proper TVD codes. Other processes like
molecular thermal dissociation and radiation emission which might occur in
such modified, radiative shocks are not in the scope of the present study. 
It is, therefore, appropriate to employ an ideal MHD code for 
investigating the problems of this paper. In such codes electrical resistivity 
is mimicked through numerical approximations,
allowing such important real MHD effects as reconnection. 

We have used the multidimensional, Cartesian version of the code 
(Ryu, Jones, \& Frank 1995) with a new ``Method of Characteristics, Constrained
Transport'' scheme for preserving the
$\nabla\cdot{\bf B} = 0$ condition as described in Ryu \etal (1998). 
The computational domain is on the $xy$ plane and both 
magnetic field and velocity $z$-components are set to zero.
The resolution characterizing all the cases listed in 
Table \ref{tabset} is 50 zones per cloud radius. 
The dimensions of the grid are given in the fifth column of the same Table. 
When exact mirror symmetry exists across the X-axis (Cases 1, 3, 4, 8 and 9), 
only 
the plane $y\ge 0$ has been included in the computational box, thus halving 
the grid size. Inflow conditions are set on the left boundary, whereas 
top and right boundaries are always open. The bottom boundary is also open 
except for Cases 1, 3, 4, 8 and 9 for which it is reflective.

Except that in one case (Case 9), in which radiative losses are being taken
into account, we generally assume an adiabatic flow with index $\gamma=5/3$ 
($p = [\gamma-1]e$). The clouds are always initially uniform and in
pressure equilibrium with the background medium. The cloud density $\rho_c$
is greater than the background $\rho_i$ by a factor $\chi=\rho_c/\rho_i$. 
We set our units so that the initial pressure $p_0=3/5$ and $\rho_i=1$,
giving the background sound speed $c_{si}\equiv(\gamma p_0/\rho)^{1/2}=1$.
The initial magnetic field is also uniform and its strength
is conveniently expressed by the parameter $\beta$ 
(see eq. \ref{beta}). We assume $\beta = 4$ in all the calculations. Since the 
thermal pressure ($p_0$) has not been specified in any physical units, 
the assumed value for $\beta$ does not imply any particular strength
of the magnetic field. However, for typical conditions in the ISM
for which $p_0/k_{\tiny B}=1600$ K cm$^{-3}$, $\beta = 4$
corresponds to $B\sim 1.2 \mu$G. 
Self-gravity has been neglected throughout our calculations. We have 
assumed very simple initial conditions for our simulations. In particular, 
several quantities, such as pressure and magnetic field, are initially
set as uniform, which is certainly not the case for the ISM. However, these
approximations are not too important for the objective of our study. In
particular the supersonic motion of the clouds is not affected by the 
initial pressure balance. Moreover,
the dynamical interaction of clouds and magnetic field (with the
deceleration of the former and the local amplification of the latter) can only
be significant for the large scale component of the magnetic field. In fact
magnetic field lines can be stretched only if their extent is much larger than 
the cloud size. On the other hand, local inhomogeneities would be quickly 
advected past the cloud causing only minor deviations. 
Thus, as long as there is an underlying ``uniform'' component to the ISM
magnetic field on the scale of a few parsecs, the presence of local
irregularities will not substantially alter the behavior expected. Interstellar
polarization observations (\eg \cite{mf70}) and Faraday rotation measures 
(\cite{gw63}, \cite{mb64}, \cite{vk75}) show clearly the existence of such an 
underlying uniform field.

With Case 9 in Table \ref{tabset} we intended to investigate the possible
effect of radiative losses on the cloud interaction with the magnetic field. 
The radiative cooling function we have used includes free-free emission, 
recombination lines, and collisional excitation lines as well as heating 
terms provided by collisional ionization and ionization by cosmic 
rays. Its properties are fully described in Ferrara \& Field (1994) whereas its
implementation is detailed in Miniati \etal (1998). Since the cooling function
is not crucial for the issues related in this paper, we refrain from
going into further details and refer to the above references for a more
complete description. 

The clouds are initially set in motion with a Mach number $M=v_c/c_{si}$,
where $c_{si}$ is the sound speed in the inter-cloud medium. As
shown in Table \ref{tabset} different Mach numbers have been employed 
in this work. In addition,
we explore several values of $\theta=\arccos[{\bf v_c\cdot B}/(v_c B)]$,
ranging from the aligned case ($\theta
=0^\circ$) to the transverse case ($\theta =90^\circ$), as well as
several values of $\chi$, as listed in Table \ref{tabset}.

\section{Results}
\label{res}

\subsection{Line Draping in a Oblique Magnetic Field}
\label{obl}

It was clear from previous numerical studies mentioned in \S \ref{desp} that
the orientation of the magnetic field lines with respect to the cloud
motion plays a key role in the interaction between the cloud and 
the magnetic field. It is important, therefore, to investigate in more
detail the general case of cloud motion making an angle 
$\theta$ with the initial uniform field (see Figure \ref{diagr}). 
Intuitively we expect that the transverse field scenario should be the 
most common one. In fact,
for any $\theta>0$, as the cloud moves through the magnetized background
and pushes the field lines at its nose, the field lines would tend to wrap 
around the cloud. We have tested this expectation by performing
numerical simulations involving various choices of $\theta$.
Some results are presented in Figure \ref{bxy3},
which shows, for the two cases $\theta=45^\circ$ and $10^\circ$
the density distribution with the magnetic field lines superimposed. Each
time corresponds to $t=13\tcs =19.5\coll$ and $15\tcs =22.5\coll$ 
($\coll=R_c/v_c= \tcs/M$) respectively. We emphasize that
if the field lines are frozen in the plasma, then it takes about 2$\coll$ 
for them to be transported across the cloud diameter by the background flow and
to start being stretched.

The first thing we note is that in both cases ($\theta=10^\circ$ and $45^\circ$)
the magnetic field lines drape around the cloud,
similarly to the transverse field case (\cite{jrt96}). However it is
clear that the magnetic field is significantly stronger for 
$\theta=45^\circ$ than for $\theta=10^\circ$. 
Even though the field lines drape around the cloud, 
in order for cloud motion to substantially amplify the magnetic field and, 
therefore, generate the magnetic shield, a significant amount of 
line stretching needs to occur. This need not take place
for all $\theta > 0$. Therefore, in the following we will attempt to identify 
more clearly the range of angles $\theta$ over which 
stretching of the field lines turns out to be efficient.
In addition, other 
factors are important. In particular the cloud could be decelerated by 
the ram pressure of the background medium or disrupted by instabilities
before the magnetic field lines are substantially stretched. These processes
are regulated by the cloud Mach number ($M$) and density contrast( $\chi$).
It is, therefore, also in the objective of this study to 
determine the conditions, in terms of $M$ and $\chi$, that allow the 
formation of the magnetic shield.  

\subsection{Magnetic \& Kinetic Energy Evolution}
\label{mke}

In order to explore these issues we performed a variety of numerical
simulations, characterized by the different parameters listed in Table
\ref{tabset}.
These simulations include: Cases 1-3 (set 1), corresponding to 
low density ($\chi=10$) but high Mach 
number ($M=10$) clouds, with 
$\theta = 0^\circ,~45^\circ,$ and $90^\circ$ respectively; 
Cases 4-9 (set 2), represent the opposite pairings of
high density ($\chi=100$), but low Mach number ($M=1.5$) clouds, with
$\theta=0^\circ,~10^\circ,~30^\circ,~45^\circ,$ $90^\circ$  and 
$90^\circ$ respectively.
Case 9 is identical to Case 8 except for being radiative. 
Finally Cases 10 and 11 (set 3) include clouds with $\chi$ and $M$ values
that are both either small 
($\chi=10,~M=1.5$) or large ($\chi=100,~M=10$) with $\theta=45^\circ$.
Figure \ref{plt1} shows the time evolution of the magnetic energy increment 
normalized to the initial cloud kinetic energy (left panels) and of the
relative kinetic energy decrement (right panels) for cases of set 1 (top), 
2 (middle) and 3 (bottom) respectively (see caption for details). 
Time is espressed in units of the sound crossing time, which only depends on
the cloud size and the background sound speed, but not on the parameters that 
differenciate the various cases listed in Table \ref{tabset}.
It is apparent that in most of the cases the magnetic energy increment 
amounts to an important fraction of the cloud kinetic energy. This is
especially true for set 2, which shows that even for $\theta=30^\circ$ that 
fraction is around 15\%.
On the other hand, the right hand plots show that the evolution of the kinetic
energy is affected by the interaction with the magnetic field. For all cases 
in set 1 and 2 clouds are respectively 
characterized by the same parameters: it appears
that the stronger the cloud interaction with the magnetic field (measured as 
the increment in the magnetic energy), the quicker the kinetic energy decrement.
In particular for Cases 6, 7, 8 and 9 of set 2 the differences in the kinetic
energy roughly correspond to differences in the magnetic energy. It is 
clear also that the presence of radiative losses is no relevant
the dynamical interaction of the cloud with the magnetic field.
We point out (and we shall justify it later) that flattening of the magnetic
energy growth is in general due to both cloud deceleration and outflow of
enhanced field regions. 

\subsection{Magnetic Field Reconnection in the Cloud Wake}
\label{mfr}

In the following subsections
we shall investigate and calculate in more detail the growth rate for the 
magnetic field strength (\S \ref{anls} and \S \ref{tms}). Before that,
we intend to explore another important issue
concerning the conditions of the magnetic field in the cloud wake. 
Recall that for $\theta =0$ a strong magnetic rope developed there in the
previous simulations. We see that even at
very small angles $\theta$, because the field lines tend to drape around the
cloud shape, the topology of the magnetic field in the cloud wake is 
basically that of the
transverse field case. As a result, the post-cloud ``flux rope'' 
only occurs in very special cases and is mostly the outcome of
the artificially exact alignement between the cloud motion and the magnetic
field lines. In all the other cases we have carried out, magnetic field lines,
converging in the cloud wake from the above and below, form an
antiparallel region separated by a thin current sheet in the wake. 
This configuration is classically unstable to resistive,
``tearing mode'' instability. In fact, the current sheet in the transition 
layer between the field lines of opposite direction eventually
becomes much thinner than its width (Biskamp 1993, p. 152 for details). 
The occurrence of the instability breaks up the transition current sheet into
line currents and several closed field line loops arise as field lines 
reconnect across the sheet (\eg Melrose 1986, p. 151). This process is clearly 
illustrated in Figure \ref{linev}, which shows a sequence of three panels
representing the evolution of the magnetic field at $t/\tcs = t/(M\coll)$ = 
12, 13 and 14
respectively, for Case 5 ($\theta=10^\circ, M=1.5, \chi=100$). In the top panel
the field lines in the transition layer appear distorted. Such distortions lead
later on to the formation of field line loops there,
as shown in the next panel corresponding to $t=13\coll$. These loops have
formed through reconnection processes and they provide the signature of the
tearing mode instability (Melrose 1986, p. 151). The overall process
responsible for the modification of the magnetic field topology is commonly
referred to as ``tearing-mode reconnection'' (\eg Melrose 1986). 
At $t=14\coll$ small field line loops still exist, although they are 
annihilating and are finally being convected away from the cloud.

It usually appears that when the magnetic field is oblique with respect to the
cloud motion, more tearing-mode reconnection activity takes place in the
cloud's wake, as compared to the exactly transverse field case. 
We attribute this property to 
the broken symmetry across the X-axis, which makes the flow in the wake 
more irregular, thus stimulating the onset of the instability. Moreover,
the uneven magnetic tension caused by the broken symmetry of the magnetic 
field perturbs further the motion of the cloud. In fact it induces a 
vortical motion of the cloud body, which affects annihilation processes of
the magnetic field inside the cloud itself. In addition, since the magnetic 
shield develops asymmetrically on the cloud nose, the cloud tends to drift 
perpendicular to its initial motion, toward the direction of lower magnetic
pressure. 

\subsection{Analysis of the Simulations}
\label{anls}

It is common to compare the magnetic energy in a cloud to its kinetic and/or
thermal energies 
(Heiles 1989, Vershuur 1989, Myers \& Khersonsky 1995, Myers \etal 1995),
in order to find insightful trends among the various physical quantities of 
the ISM. Moreover, if the generation of local irregularities in the
interstellar magnetic field is in part related to the motion of dense clouds
(see \S \ref{motiv}), 
and in particular is generated at the expense of their kinetic energy, then
such comparison may turn out to be relevant for
the overall energy budget of galaxies. 

The equation describing the evolution of the magnetic energy $E_B$ 
in a volume $V$ due to field line stretching is (\cite{ch61})
\begin{equation}
\frac{\partial E_B}{\partial t} = \frac{1}{4\pi}
\int_V B_i\frac{\partial v_i}{\partial x_j}B_j dV 
\sim 2 \frac{v_c}{R_c}\sin\theta E_B = 2 \frac{E_B}{\coll}\sin\theta,
\label{em1}
\end{equation}
where $\sin\theta$ takes into account the field
orientation. This  generalizes eq. 10 in Jones \etal (1996). 
If $E_B$ and $E_{B0}$ are the 
magnetic energy in the region where the stretching of the field lines takes 
place at an arbitrary time $t$ and at $t=0$
respectively, then an approximate
solution to equation \ref{em1} is:
\begin{equation}
E_B=E_{B0}~e^{2t\sin\theta/\coll}.
\label{sem1}
\end{equation}
Equation \ref{sem1} describes an exponential growth of the 
magnetic energy with a timescale as short as $\coll/\sin\theta$. 
This approximation holds true only for the very beginning of the
evolution. Depending on initial field strength, nonlinear effects
(\ie from flow modifications) soon become
important so that a different approach must be employed. Starting from eq. 
\ref{em1},
we assume that the magnetic energy growth is initially exponential. 
A simple way to modify this for nonlinear growth is to begin from
the representation of the exponential function as a series of power laws, each 
with its appropriate coefficient. We may express the saturation of
field growth by saying that the back reaction of the magnetic
field suppresses the contribution of high order terms that dominate
the exponential function for large arguments.
On the other hand, our simulation data show that the relative magnetic energy 
enhancement ($\Delta E_{B}/E_{B0}$) is 
faster than implied by a first order approximation (linear growth with time),
but is well described by a power law with index $2\ge m \ge 1$ (see the next
paragraph).
The physical meaning of a power law behavior can, at least in part,
be inferred from eq. \ref{em1}. Specifically it implies that 
the velocity shear $\partial v_x/
\partial y$, which in eq. \ref{em1} is represented as a constant 
$\sim v_c/R_c$, actually decreases with time like $\sim t^{-1}$ once the 
magnetic field begins to modify the flow.
That behavior makes intuitive sense if one considers that
the magnetic field affects the direction of the gas flow around the 
cloud before the cloud velocity itself is decreased. Thus, the effect is 
probably mostly due to the growing 
size of the magnetic shield, the region where field line stretching occurs, and
an associated broadening of the shear layer around the cloud.
That is, the scale represented by $R_c$ in eq. \ref{em1} apparently
increases linearly with time.
We derive an estimate of the magnetic energy growth
by expanding the exponential function in eq. \ref{sem1} and retaining 
a single (presumed dominant) term of the power law series, with index $m$ to be determined:
\begin{equation}
\frac{\Delta E_B}{E_{B0}} \equiv  \frac{E_B}{E_{B0}}- 1 =  
\left(e^{2t\sin\theta/\coll} - 1 \right)
\sim \left(\frac{2tM\sin\theta}{\tau_{cs}}\right)^m \frac{1}{m!},
\label{ener1}
\end{equation}
where in the last expression we used $\coll=\tcs/M$.

Eq. \ref{ener1} was tested for all cases listed in Table \ref{tabset}.
We plotted $\log(\Delta E_B/E_{B0})$ versus
$\log(t/\tau_{cs})$. If these two quantities are related as described by 
eq. \ref{ener1}, then the plots should resemble
\begin{eqnarray} \nonumber
\log\left(\frac{E_B}{E_{B0}}-1\right)&\simeq &\log\frac{2^m }{m!}
+m\log M+m\log(\sin\theta) +m\log\frac{t}{\tau_{cs}} \\ &=&
A(m,M,\theta)+ m \log\frac{t}{\tau_{cs}}
\label{lener1}
\end{eqnarray}
and their slope should correspond to the index $m$. 
Log plots of the magnetic energy increment (this time normalized to the initial
magnetic energy) as a function of time are shown in Figure \ref{plt2}. Left 
plots are log analogous of those in Figure \ref{plt1} (left panels) and 
refer to the magnetic energy over the whole grid.
On the other hand, right plots correspond to the magnetic
energy in a region of size $R_c \times 2R_c$ and placed before the
cloud's nose. As we can see, the growth of the magnetic field is much stronger
in the latter case than in the former, so that the magnetic energy density is
much higher at the cloud nose than anywhere else. However, because that is 
only a small region, its contribution to the total increase in the magnetic
energy is not dominant. In other words, most of the exchange of kinetic and 
magnetic energy takes place on a more extended region. 

In general, in both cases the power law description of the magnetic energy 
growth with time applies only for $\theta\ge 30^\circ$. So, we will not 
consider cases with $\theta < 30^\circ$ any further.
Strictly speaking, only cases belonging to set 2 (and with $\theta 
\ge 30^\circ$) follow the power law behavior. Curves corresponding to 
all other cases exhibit a turn over at some point in their evolution. 
That is particularly evident in set 1. There are two main reasons for that
behavior and both are related to the fact that 
cases of sets 1 and 3 are characterized by high Mach numbers ($M$) and/or 
low contrast density ($\chi$). In fact such parameters determine a quick
deceleration of the cloud by the background ram pressure (\S \ref{desp}). 
On the one hand, that
reduces the time available for the cloud to stretch the field lines 
(\S \ref{tms}), therefore stopping the growth of the magnetic energy. On the 
other hand, since the cloud approaches more quickly the computational
boundary, outflow of magnetic energy becomes soon important and significantly
affects the growth curve of the magnetic energy. 

Recognizing these limitations we can still calculate values of the slope ($m$)
and the intercept ($A$) for all cases with $\theta \ge 30^\circ$, restricting 
ourselves to the growing portions of the curves for set 1 and 3. 
The last four columns of Table \ref{tabset}
report values of the slope as well as of the intercept for the curves in Figure
\ref{plt2} of the 
magnetic energy in the total grid ($m,A$) and in the ``cloud nose region''
($m_n,A_n$) respectively, as obtained 
with a simple least $chi-square$ method (\cite{tay82}). For cases of set 
1 and 3 the resulting slopes are somewhat sensitive to the choice of the 
``growing portion'' of the curves. 
We point out that we report these values with the only purpose of 
giving a qualitative description of the evolution of the quantities considered;
therefore they should not be taken as to indicate anything too precise.
Values of $m$ and $m_n$ given in table \ref{tabset} 
average around $\bar{m}=1.43$ and $\bar{m}_n=0.9$  respectively. 
In the following we shall simply take the representative values,
$m\simeq 1.5$ and $m_n\simeq 1.$. 
Finally, it is worth noting that for $\theta\ge 30^\circ$ 
the distance between the different curves (given by the difference
in their $A$ coefficients) is
consistent with what expected from eq. \ref{ener1} for different values of
$M$ and $\theta$.
This further supports the approach of eq. \ref{ener1}.

\subsection{Dimensionality Issues}
\label{dis}

In order to validate our 2D MHD numerical calculations, it is important to
address the limits imposed by the symmetry along the cylinder axis due to the
2D geometry. For that purpose we carried out an analogous 3D version of Case 8,
simulating the supersonic motion ($M=1.5$) of a dense ($\chi=100$), finite
cylinder through a transverse magnetic field ($\theta=90^\circ$). 
The length of the cylinder semi-axis was $\ell_z=3R_c$, where $R_c$ was its
radius.  Motion was perpendicular to the cloud axis. Exploiting the mirror 
symmetry
across the planes $z=0$ and $y=0$, we were able to limit the calculation to
only one quarter of the total actual volume. Therefore  
we used a computational domain $x$=[0,20], $y$=[0,8] and $z$=[0,8] with the
cloud center initially located at $x_c=2R_c$, $y_c=z_c=0$. The resolution was
16 zones across a cloud radius with $R_c=1$ in numerical units.
In order to make a sensible comparison with Case 8 we calculated 
the growth of the magnetic energy in the plane $z=0$.
In fact, if the 2D approximation is realistic, then the evolution of the
magnetic energy on the plane $z=0$ resulting from the 3D calculation should
resemble that seen in Case 8. 
The results from the 3D calculation are shown in Figure \ref{3Dme}. In the
three panels we
report: log plots of the relative increment of the magnetic energy
on the whole grid (top) and in the usual region of 
size $R_c\times 2R_c$ before the cloud nose (middle)
versus time expressed in units of the sound crossing time .
Also we report the evolution of the relative change of 
the kinetic energy (bottom). The
Figure shows that up to about one $\tcr$, the growth of the magnetic energy is
qualitatively the same as in the 2D calculations. Following the same procedure
as in \S \ref{anls}, for this case we infer
$(m,A)=(1.26,1.3)$ and $(M_n,A_n)=(0.8,0.6)$. 
This slope is slightly smaller than in 
the previous case, meaning that in 3D, the growth of the magnetic 
field is somewhat slower, as expected, but is still faster than linear with
time. 

\subsection{Timescale}
\label{tms}

Given the slope $ 1\lesssim m \lesssim 1.5$, 
in this section we derive the timescale 
$\tau_{ma}$, to transform half of the cloud kinetic energy into 
magnetic form (formation of the magnetic shield).
The cloud initial kinetic energy, $E_{k0}$,
can be expressed in terms of $E_{B0}$, the initial magnetic energy in the 
region of volume $V_s$ where field line stretching occurs, as 
\begin{equation}
E_{k0}= \frac{1}{2}\gamma \chi M^2\beta ~\frac{V_c}{V_s}~E_{B0}
\label{kin0}
\end{equation}
where $V_c$ is the volume of the cloud. 
Thus, setting the magnetic energy at $t=\tau_{ma}$ equal to half the initial 
cloud kinetic energy, and considering eq. \ref{ener1} we obtain
\begin{equation}
\left(\frac{1}{4}\gamma \chi M^2\beta\frac{V_c}{V_s} - 1 \right)
 \simeq  \frac{1}{4}\gamma \chi M^2\beta\frac{V_c}{V_s}
 = \left(\frac{2\tau_{ma}\sin\theta}{\coll}\right)^{m} \frac{1}{m!}\simeq
 2\left(\frac{\tau_{ma}\sin\theta}{\tau_{cs}}\right)^{m}M^{m};
\label{time1}
\end{equation}
(the numerical factor in the last expression derives from the assumption 
$m\sim 1$). From eq. \ref{time1} we derive
\begin{equation}
\tau_{ma} = \left(
\frac{\gamma \beta\chi}{8}\right)^\frac{1}{m} 
\frac{M^{\frac{2}{m}-1}}{\sin\theta}\left(\frac{V_c}{V_s}\right)^\frac{1}{m}
~\tcs 
\end{equation}
and for $m=1.5$
\begin{equation}
\tau_{ma} =
\beta^{2/3} M^{1/3}
\left(\frac{\chi}{100}\right)^{2/3}
\left(\frac{R_c}{0.5\mbox{pc}}\right)
\left(\frac{10\mbox{km sec}^{-1}}{\ssp}\right)~5.3\times 10^5 \mbox{yr}
\label{tauma} 
\end{equation}
using a representative $\sin\theta\sim \sin45^\circ =1/\sqrt{2}$ and neglecting
the factor $(V_c/V_s)^{2/3}$ which probably amounts to a few tenths.
This is the timescale over which a cloud gives up half of its kinetic energy 
to the background magnetic field by stretching the field lines.
When viewed the other way around, $\tau_{ma}$
is also {\it the timescale over which a cloud embedded in a magnetized flow
is accelerated to move with the background plasma through magnetic forces}. 
Therefore, $\tau_{ma}$ can be considered a ``magnetic acceleration'' timescale.
We emphasize again such description is valid only up to several $\tcr$.

\subsection{Physical Issues}
\label{phi}

We can now compare $\tma$ with the drag timescale $\tdr$ 
(eq. \ref{eqdr}),
over which a cloud is stopped by the ram pressure (\S \ref{desp}).
Their ratio (using $\gamma=5/3,~\theta=45^\circ$ and $\beta\sim$ a few) is: 
\begin{equation}
\frac{\tma}{\tdr}\sim \chi^{\frac{1}{m}-1}M^{2/m}.
\label{macr}
\end{equation}
Since $m > 1$, 
Eq. \ref{macr} shows that for small Mach numbers ($M$) and large density 
contrasts ($\chi$) the magnetic tension is more important than the drag force.
For example, for a ``standard'' interstellar cloud characterized by 
$\chi\sim 100$ and 
$M\sim 1.5$ (if $m=1.5$) $\tau_{ma}/\tau_{dr}\sim 0.37$, so that 
a significant fraction of the cloud kinetic energy can be transformed into 
magnetic form. We point out that we use the drag time here instead, for
example, of the crushing time because as long as the cloud is moving 
the magnetic field will be enhanced by stretching even after the passage of the
crushing shock. Table \ref{tabts} summarizes the relevant timescales
characterizing the dynamical eveolution of a typical cloud in motion.

In Figure \ref{cutm} we report cuts of the log of the magnetic energy density 
($B^2/2$, in numerical units)
along the axis of motion of the cloud for the following three representative
cases: Case 3 (solid line), Case
7 (dotted line) and Case 11 (dashed line).
The time corresponding to each of these cuts is
expressed in terms of various dynamical timescales in Tab. \ref{tcom}.
Values in column 6 show that the evolutionary time is still below several
$\tcr$ so that instabilities in the third dimension have not become too 
disruptive yet.
In the same table we also calculate the ratio in eq. \ref{macr} for each case.

The three lines in Figure \ref{cutm} clearly exhibit a peak corresponding 
to the magnetic shield 
(\S \ref{desp}). For Case 3, the magnetic energy density 
$B^2/2\sim 40$ is comparable with the ram pressure of the background flow
($\sim \roi v_c^2 \sim 100$), in accord with Jones \etal (1996). For this case
 $\tma/\tdr \sim 9 \gg 1$. Therefore, the cloud is mostly being decelerated 
by the ram presure of the flow and soon after, 
as the magnetic field lines relax, the magnetic energy drops. 
This effect, along with magnetic energy outflow through the boundary, is
responsible for the turn-over visible in Figure \ref{plt1} for Cases 2 and 3. 
Similarly, for Case 11 (dashed line), 
since $\tma\gg\tdr$, only a small fraction of the
cloud kinetic energy is transformed into magnetic energy. However, because
of the large mass of the cloud ($\chi\gg 1$), in this case even a small 
fraction of the cloud kinetic energy is enough to amplify the magnetic field
in the magnetic shield beyond the background ram pressure limit.
Finally, in Case 7 (dotted line), $\tma/\tdr \sim 0.5$ 
and, as we can see, $B^2/2~(\sim 12.5$) is much larger than 
the ram pressure limit ($p_{\rm ram} = \rho v^2 \sim 2.25$). 
The magnetic energy 
is still a small fraction ($\sim 20$ \%, see Fig. \ref{plt1}) 
of the initial cloud kinetic 
energy, however. This is roughly in accord with the model
\footnote{Acording to eq. \ref{ener1} (with $m=1.5$),
after a time $t/\tma\sim 0.42$ only a fraction
$\sim \slantfrac{1}{2}\times (0.42)^{1.5}\sim 0.14$ of the {\it total} initial 
kinetic energy is converted into magnetic form.}. 

We shall now return to the issue raised in \S \ref{desp} about the limit
on the magnetic energy that can be generated in the magnetic shield.
It is clear that as long as the lines are being stretched by the gas
motion, then the magnetic field energy will increase. This result was tested
with a numerical simulation of a very large density ($\chi=10^3$) 2D cloud 
with jet-like shape (in fact a sheet of dense gas). This object had a very
large kinetic energy in its motion with respect to the background gas
and it turned out that the magnetic energy grows (almost
linearly) until it becomes so large on the cloud nose that the code
cannot handle the calculation properly anymore. However, by contrast, it 
is usually observed in the solar system that there is a 
sort of ``equilibrium'' between ram pressure and magnetic pressure around
massive objects moving supersonically with respect to the solar wind.
This is the case for example at the earth magnetopause (\eg \cite{parks91})
and also found in simulated interations between the solar wind and
outflows from comets (Yi \etal 1996). That different result is due to
the fact that in such cases the physical system reaches a stationary
state, whereas the clouds considered here are never in a stationary
flow. It is easy then to show that in a steady flow  the previous limiting
relation between magnetic pressure and ram pressure must hold. It is important 
to observe that stationarity is achieved in the aforementioned solar system
cases because the basic motions of those objects are not affected by the
flow and because the structural changes brought about through their
interactions with the solar wind are strongly resisted. In fact the magnetopause is
supported by a dipole magnetic field which responds strongly to any attempt of
deformation ($B^2 \propto R^{-6}$). Also the comet outflow provides a
momentum flux (dynamic pressure) that varies as the inverse square of the
distance ($\propto R^{-2}$). So again it is difficult to deform the
shape of this kind of object. These facts allow the magnetic field lines
to easily slip around the surface for a ``quick'' attainment of stationary 
state. In our 2D cloud simulations the
field lines are confined in the plane, so that they are bound to be stretched
as long as the cloud moves. Growth is ultimately limited by deceleration
of the clouds. However, even in 3D calculations since clouds
undergo strong deformations on their leading edge the stationary state
is never reached and magnetic flux can become trapped. In fact
both the magnetic field and the cloud shape change continously in response to
each other dynamical effects. Even in 3D MHD cloud simulations it turns out 
that the magnetic pressure in the shield becomes larger than the ram pressure
by a factor of at least a few (\cite{gmrj98}).

With this perspective, it can be concluded then that an accurate estimate of
the upper limit for the magnetic energy in the magnetic shield can only come
from appropriate 3D numerical simulations. However, it is immediately clear
from the investigation presented in this paper that previous studies have
underestimated the extent to which the magnetic field can be enhanced by
stream motions in the ISM and overlooked the dynamical importance that this
process can have in terms of the evolution of these streams as well as
the energetics of the global ISM.
In addition, as it will be properly addressed in the following section, this
amplification of the magnetic field originating from the cloud motion
has interesting observational implications.

\section{Comparison with Observations}
\label{motiv}

We attempt now to link our results to some recently found, interesting features
of the magnetic field in diffuse clouds.
As pointed out in \S \ref{code}, the conditions in 
the ISM are such that ideal MHD is applicable even to \ion{H}{1}
clouds characterized by the scales of interest here.

Myers \etal (1995) compared data of 21 cm Zeeman effect obtained
with the 26 m radio telescope at Hat Creek Radio Observatory (HCRO) and those
with the 100 m Effelsberg telescope. 
They found,
even though the magnetic strengths obtained with the two telescopes
are usually consistent within experimental errors, that there are cases where
the field strengths are 
apparently discrepant. For instance, in the particular direction ($l,~b$)=
(141.1, 38.8), the HCRO gives
$B_{\parallel}=18.9~ \pm ~1.8~\mu $G, whereas the 100 m measurement 
is 3.5$~\pm ~3.7~\mu$G. Those authors argued that this is probably real and due
to the significant variation of the magnetic field strength inside the larger
HCRO beam compared to that in the 100 m beam.
This means that, in that region, the field has a significant structure on a 
length scale L $\le~ 1.6$pc and measuraments of $B_\parallel\ge 18.9\mu$G 
should be found by the 100 m telescope somewhere within the HCRO beam.
In \S \ref{desp} we have seen that a cloud propagation through a 
magnetized background is able to modify significantly the initially uniform
field structure. This is particularly true for $\theta\ge 30^\circ$. 
As already seen in Figure \ref{cutm}, after a time of the order of $\tcr$,
the intensity of the field in the magnetic shield can be several
times as large as the background value. 
The magnetic shield dimensions are of the order of 
the cloud radius ($R_c$). The variation in the 
magnetic field measured by Myers \etal (1995) (from 3.5~$\pm ~3.7~\mu$G up to
values $\ge$ 18.9) is consistent with the
amplification resulting from our numerical study.
Therefore, it is plausible to explain the local inhomogeneities observed 
in Myers \etal (1995) in terms of the interaction between 
the cloud motion and the background magnetic field. 

Another interesting observational finding is that in many \ion{H}{1} clouds,
the magnetic and kinetic energy densities have comparable values
(Heiles 1989, Vershuur 1989, Myers \& Khersonsky 1995, Myers \etal 1995). 
For some cases the magnetic energy is actually in excess of the kinetic 
(Myers \etal 1995). 
Myers \etal (1995) point out that if collisions between neutral and ions are 
high enough, MHD waves will propagate 
through the field-fluid system at the Alfv\'en speed based on the combined
neutral/ionic mass density.
The kinetic and magnetic energy densities in an Alfv\'en wave
are in equipartition.
Therefore, if the magnetic energy in the waves 
is similar to that in the mean magnetic field, 
the nonthermal line 
broadening would correspond to the Alfv\'en speed for the neutral/ionic
mass.
In other words, the measured magnetic and kinetic energies of the gas 
should be similar.
On the other hand, those observational results are in agreement with the 
depiction given in this paper, according to which
the magnetic energy is enhanced by field lines stretching
produced by coherent gas flow, \ie by cloud motions.
Since this field enhancement occurs at the expense of the cloud kinetic energy,
it makes sense that some equipartition between magnetic and kinetic energy
is achieved. Eventually the cloud is stopped, either by the magnetic field
tension or by the ram pressure of the background flow, so that at some 
point the kinetic energy becomes even smaller than the magnetic energy 
(although the latter must still be less than the {\it initial}
cloud kinetic energy). On the other hand we have seen that only a 
weak amplification of the magnetic field occurs, when the latter is mainly 
aligned with the cloud motion.
Therefore, not for all clouds magnetic and kinetic energy are expected to be 
comparable.

\section{Conclusions \& Summary}
\label{cs}

We have studied the motion of an individual 2D clouds through a magnetized 
lower density background medium, for different density contrasts, $\chi$,
Mach numbers, $M$, and field geometries, $\theta$. 
Our major findings can be summarized in the following items:
\begin{itemize}
\item A cloud moving through a magnetized medium tends to 
interact with the magnetic field in a way that depends upon the angle 
$\theta$ between the initial cloud velocity and background magnetic field.
According to our simulations, for large angles $\theta$ the magnetic field 
lines are stretched efficiently by the cloud motion and, therefore, the 
magnetic field is significantly amplified at the expense of the cloud 
kinetic energy. This is demonstrated here for angles $\theta\ge 30^\circ$. 
Our simulations then suggest that when
$\theta\ge 30^\circ$ the magnetic shield developed by a moving cloud
may behave as an efficient ``bumper'' in MHD cloud collisions 
(\cite{mrfj98}).
\item Our numerical experiments show a global magnetic energy growth
reasonably well described by a power law in time with index $m\simeq$ 1.2--1.5
for a wide range of values of density contrast ($\chi$), Mach number ($M$)
and field geometry ($\theta$). This behavior is not significantly affected by
the presence of radiative losses. In addition, it seems confirmed by at least 
one analogous 3D simulation as well ($m\sim 1.2, 1.3$). 
We define a characteristic timescale 
$\tau_{ma} \sim (\beta\chi)^\frac{1}{m} M^\frac{2-m}{m}~
\frac{\tcs}{\sin\theta} $
for part of the cloud kinetic energy to be transformed into magnetic energy.
We note, as well, that $\tau_{ma}$ can also be interpreted as the time for 
a cloud to be accelerated to the flow velocity by Maxwell stresses.
Since $m >1$, for high $\chi$, low Mach number clouds this dynamical timescale
would be shorter and therefore more relevant then the drag timescale
$\tdr\sim \chi^\frac{m-1}{m}M^{-\frac{2}{m}}~\tma$.
\item We suggest that this mechanism, if confirmed more generally by 3D MHD 
simulations, could provide an explanation for some characteristics observed 
in the interstellar magnetic field. In particular the existence of magnetic
field inhomogeneities on a scale about 1 pc in correspondence of 
\ion{H}{1} clouds can be interpreted as the formation of the magnetic shield,
as outlined in \S \ref{motiv}. In addition the equipartition of kinetic 
and magnetic energy, independent of whether or not the cloud is 
self-gravitating, would follow naturally from this depiction where 
the the magnetic energy is provided by the cloud kinetic energy itself.
\item In the majority of the cases we simulated the magnetic field lines drape around 
the propagating cloud, even though no strong magnetic shield forms. 
In addition, as the symmetry across the motion axis is broken (for
$0^\circ<\theta <90^\circ$), the flow
in the wake becomes more turbulent, thus facilitating the onset of
tearing mode instabilities which weaken the magnetic field strength there.
This means that the ``flux rope'' mentioned by earlier authors might 
develop only in very special cases or exist only temporarily.
Asymmetries in the magnetic tension around the cloud can induce
vortical motions within the cloud itself. Those motions, in turn, could influence 
its subsequent development or annihilation of fields within the cloud.

\end{itemize}

\acknowledgments

We are grateful to Ruth Dgani, the referee, for insightful comments and
interesting discussions which improved the manuscript. 
We acknowledge the support provided by Gianluca Gregori in performing the 
3D MHD numerical calculation presented in this paper.
The work by FM and TWJ was supported in part by the NSF through
grants AST-9619438, INT-9511654, by Nasa grant NAG5-5055 and
by the University of Minnesota Supercomputer Institute.
The work by DR was supported in part by KOSEF through grants
975-0200-006-2 and 981-0203-011-2.

\clearpage

\begin{deluxetable}{cccccccccc}
\footnotesize
\tablecolumns{10}
\tablecaption{Summary of 2D-MHD Single Cloud Simulations \label{tabset}}
\tablehead{  \multicolumn{10}{c}{$\beta$= 4, $\gamma$=5/3}  \\
\cline{1-10}\\
\colhead{CASE \tablenotemark{a}} &
\colhead{$\chi$} &
\colhead{M\tablenotemark{b}} &
\colhead{$\theta$} &
\colhead{Grid-Size\tablenotemark{c}} &
\colhead{End Time\tablenotemark{d}}&
\colhead{$m$\tablenotemark{e}} &
\colhead{$A$\tablenotemark{e}} &
\colhead{$m_n$\tablenotemark{f}} &
\colhead{$A_n$\tablenotemark{f}} 
\\
\colhead{}&\colhead{}&\colhead{}&\colhead{($^\circ$)}&\colhead{($R_c^2$)}&
\colhead{($\tau_{cs}$)}&\colhead{}&\colhead{}
}
\startdata
1 & 10  &  10 &  0 & 10 $\times$ 5    & 4.2 &  -   & -     &  -  & - \nl
2 & 10  &  10 & 45 & 20 $\times$ 10   & 6.6 & 1.4  & 0.2  & 0.8 & 1.6 \nl
3 & 10  &  10 & 90 & 10 $\times$ 5    & 4.2 & 1.5  & 0.7  & 0.6 & 1.8 \nl
4 & 100 & 1.5 &  0 & 15 $\times$ 15   & 13  &  -   &  -    &  - & - \nl
5 & 100 & 1.5 & 10 & 20 $\times$ 10   & 15  &  -   &  -    &  - & - \nl
6 & 100 & 1.5 & 30 & 20 $\times$ 10   & 15  & 1.4 & -1.4 & 1.1  & 0.1\nl
7 & 100 & 1.5 & 45 & 20 $\times$ 10   & 13  & 1.4 & -1.2 & 1.0  & 0.4\nl
8 & 100 & 1.5 & 90 & 20 $\times$ 7.5  & 9   & 1.5 & -1.3 & 1.0  & 0.6\nl
9 & 100 & 1.5 & 90 & 20 $\times$ 7.5  & 9   & 1.4 & -1.2 & 0.9  & 0.7\nl
10& 10  & 1.5 & 45 & 20 $\times$ 10   & 9.6  & 1.2 & -1.3 & 0.6  & 0.3\nl
11& 100 & 10  & 45 & 20 $\times$ 10   & 9.6 & 1.6 &  0.3 & 1.2  & 1.8\nl
\enddata
\tablenotetext{a}{All models use $\beta=4$, $\gamma$ = 5/3.}

\tablenotetext{b}{The Mach number is referred to the intercloud sound speed, 
$c_{si}$.}

\tablenotetext{c}{The grid size is expressed in units of cloud radius. One cloud
radius has $R_c$=50 zones.}

\tablenotetext{d}{The end time is expressed in terms of sound crossing times
$\tau_{cs}$.}

\tablenotetext{e}{Coefficients for the growth of the magnetic energy over the
whole grid. For cases 2, 3, 9 and 10 only the growing portion of the 
magnetic energy growth curves has been used to determine $m$ and $A$ (see
the text for details). These are not provided for Cases 1, 4 and 5
because the model of \S~\ref{anls} is not appropriate for these cases.}

\tablenotetext{f}{Coefficients for the growth of the magnetic energy inside a
region of size $R_c \times 2R_c$, placed before the cloud's nose. 
The same precautions as in the previous note apply here.}

\end{deluxetable}

\begin{deluxetable}{ccccccc}
\footnotesize
\tablecolumns{7}
\tablecaption{Timescales \label{tabts}}
\tablehead{  
\colhead{CASE} &
\colhead{$\chi$} &
\colhead{M} &
\colhead{$\tau_c/\tcs$} &
\colhead{$\tau_{cr}/\tcs $} &
\colhead{$(\tau_{ma}/ \tcs)  \sin\theta $\tablenotemark{a}} &
\colhead{$\tau_{dr}/\tcs $} 
}

\startdata
1,2,3       & 10  & 10  & 0.1   & 0.63 & 8.8  & 1    \nl
4,5,6,7,8,9 & 100 & 1.5 & 0.67  & 13.3 & 21.8 & 66.7 \nl
10          & 10  & 1.5 & 0.67  & 4.2  & 4.7  & 6.67 \nl
11          & 100 & 10  & 0.1   &  2   & 41   & 10

\enddata
\tablenotetext{a}{The factor $\sin\theta$ is to take into account the
slight difference between cases characterized by the same $\chi$ and $M$, but 
different angles $\theta$. For Cases 1, 4 and 5 ($\theta < 30^\circ$), as 
already mentioned in \S \ref{anls}, $\tma$ is meaningless.}

\end{deluxetable}

\begin{deluxetable}{cccccc}
\footnotesize
\tablecolumns{6}
\tablecaption{Cloud Timescales \label{tcom}}
\tablehead{
\colhead{CASE} &
\colhead{$\tma / \tdr$} &
\colhead{$t/\tma$} &
\colhead{$t/\tdr$}&
\colhead{$t/\coll$} &
\colhead{$t /\tcr$} 
}
\startdata
3 & 8.8 & 0.3 & 2.4 & 24 & 3.8 \nl
7 & 0.46 & 0.42 & 0.19 & 19.5 & 0.95\nl
11 & 5.8 & 0.14  & 0.84 & 84   & 4.2
\enddata

\end{deluxetable}

\clearpage

\begin{center}
{\bf FIGURE CAPTIONS}
\end{center}

\figcaption[]{Diagram of a cloud moving through a uniformly
magnetized background
medium with a velocity making an angle $\theta$ with the field lines.
\label{diagr}}

\figcaption[]{Density distribution with superimposed magnetic field lines 
for a single cloud propagating through a magnetized medium.
Field lines are contours of the magnetic flux.
Density figures are inverted grayscale images  of 
$\log(\rho)$.
In both cases $\beta =4$, $\chi = 100$ and $M= 1.5$.
Top panel, for $\theta=45^{\circ}$ and $t=13 \tcs = 19.5\coll$,
exhibits the formation of a 
consistent magnetic shield. Bottom panel is
$\theta=10^{\circ}$ and $t=15 \tcs = 22.5\coll$; the field lines 
are draping around the cloud, 
although the magnetic shield is not as developed as in the previous case. 
\label{bxy3}}

\figcaption[]{Time evolution of the magnetic energy 
enhancement ($\Delta E_B/E_{K0}$, left panels) and of the kinetic 
energy ($E_K/E_{K0}$, right panels), normalized to the initial cloud kinetic 
energy, $E_{K0}$,
for individual clouds propagating through a magnetized medium.
Time is espressed in units of the background sound crossing time 
($\tcs =R_c/c_{si}$) for all cases. Time labels, however,
differ because the timescales for the evolution of magnetic and kinetic energy
strongly depend on the parameters ($\chi$ and $M$) defining the 
various cases (see \S \ref{desp} and \S \ref{tms} for details).
In the top panels (set 1) open triangles correspond to Case 1, open circles to 
Case
2 and filled circles to Case 3 respectively. In the middle panels (set 2) 
open squares represent Case 4, filled squares Case 5, open triangles Case 6, open
circles Case 7, filled circles Case 8 and filled triangles Case 9. In the
bottom panels open and filled circles correspond to Case 10 and 11
respectively. The middle, left panel shows that for Case 4 ($\theta=0^\circ$,
open squares) the magnetic energy growth is slightly larger than for 
Case 5 ($\theta=10^\circ$, filled squares). 
This is only
due to the choice of the normalization parameter $E_{K0}$ which, due to 
the setting of the grid, is twice as large for Case 5 than for Case 4. 
Compare with Figure \ref{plt2} for the relative growth of the magnetic energy
normalized to its initial value. Finally, beware that
the sudden drop in the kinetic energy for Case 9 (filled
triangle) is a numerical rather than a physical effect.
\label{plt1}}

\figcaption[]{Magnetic field geometry for Case 5 ($\chi= 100, M=1.5, \theta
=10^\circ$) at three different times corresponding,
from top to bottom, to: $t/\tcs =t/(M\coll)$ = 12, 13, 14.
As in Fig. \ref{bxy3}, field lines are contours of the magnetic flux function.
Finer contouring is used, however, with respect to 
Fig. \ref{bxy3}. The sequence of panels displays the evolution of the 
``tearing-mode''
instability in the cloud wake, with the formation of closed field line loops
through reconnection and their subsequent annihilation. 
\label{linev}}

\figcaption[]{Time evolution of the relative magnetic energy 
enhancement ($\Delta E_B/E_{B0}$) relative to the whole numerical
grid (left panels) and to a region of size $R_c\times 2R_c$ 
placed before the cloud nose (right panels). In each case $E_{B0}$ is the 
initial magnetic energy relative to the region considered. Curves correspond 
to log plots of $\Delta E_B/E_{B0}$ versus time normalized to the sound crossing
time $\tcs$. As in Figure \ref{plt1}, time labels change for the
various cases, according to the different characteristic timescales.
Each case of Table \ref{tabset} is represented in the same panel and with the
same type of point as in Figure \ref{plt1}. Note that for a better
representation of Case 2 and 3 the less interesting Case 1 was not plotted in
the top, left panel. 
\label{plt2}}

\figcaption[]{3D analogous of Case 8 in
Table \ref{tabset}. For this case $\chi=100, ~M=1.5$ and $\theta=90^\circ$. 
Top and middle panels correspond to the filled circle
curves in the middle, left and right
panels of Figure \ref{plt2} respectively. Bottom panel, on the other hand,
corresponds to the filled circle plot of the middle, right panel of Figure
\ref{plt1}. 
\label{3Dme}}

\figcaption[] {Cuts through the cloud and along the cloud axis of motion,
of the log of the magnetic energy density ($B^2/2$) expressed in numerical units. 
Solid line corresponds to Case 3, dotted line to
Case 7 and dashed line to Case 10. Times corresponding to each curve and 
expressed in different units are reported in tab. \ref{tcom}. In all three 
cases it is evident a peak in the magnetic energy density due to the 
presence of the magnetic shield in front of the cloud. 
\label{cutm}}

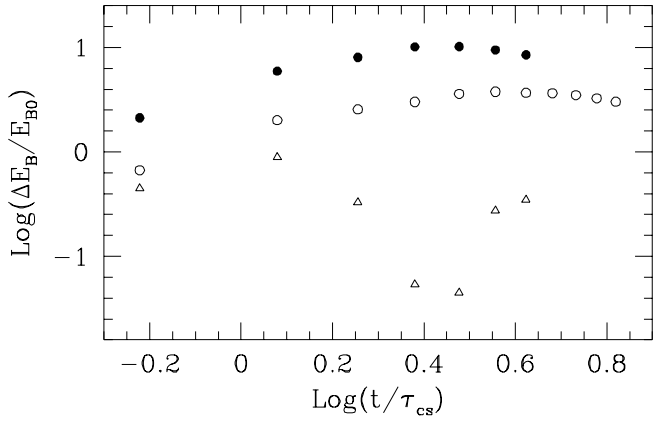
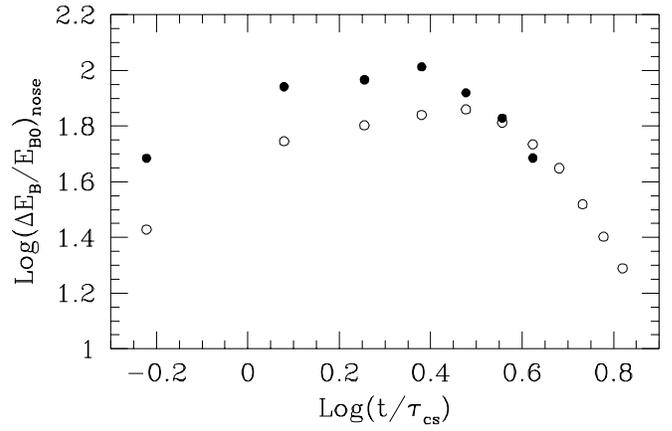
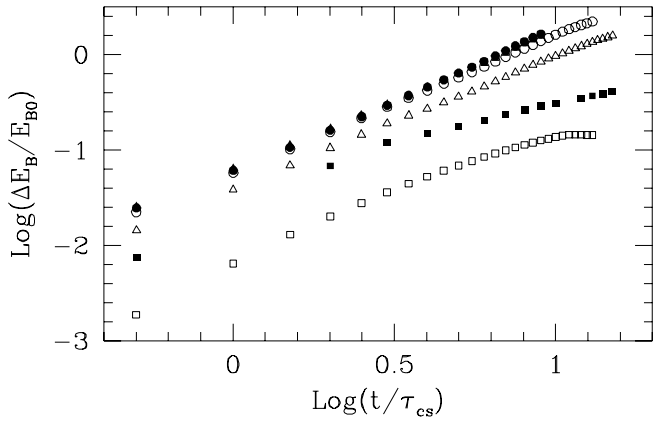
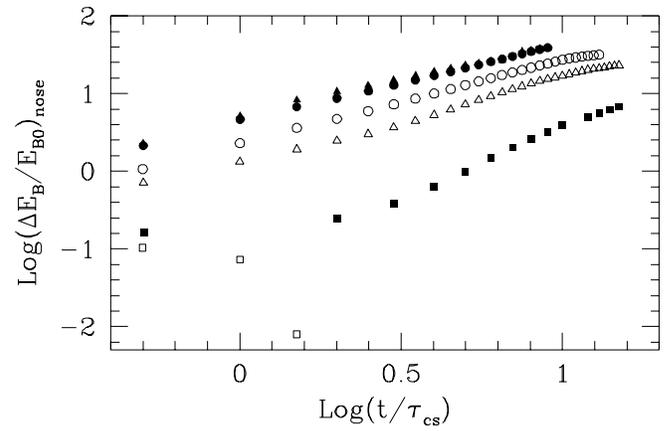
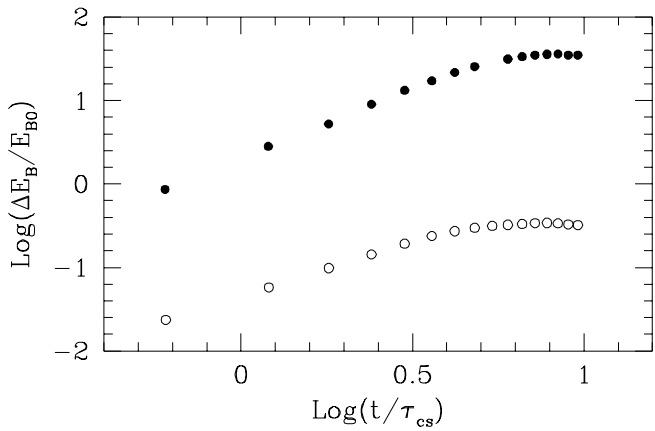
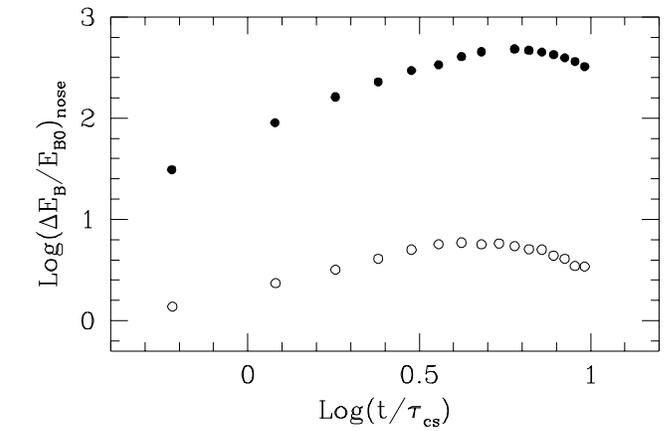
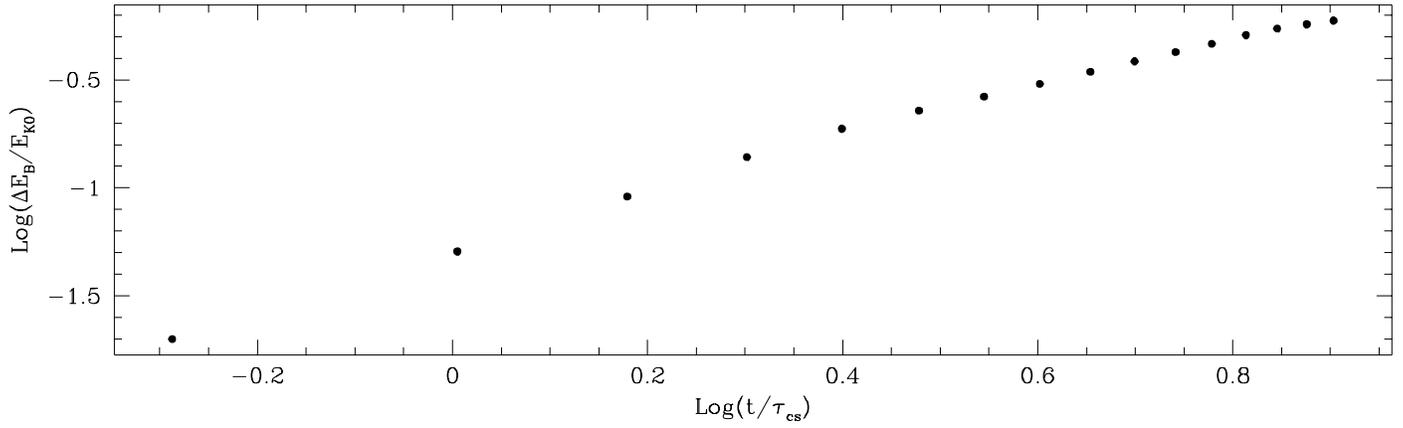
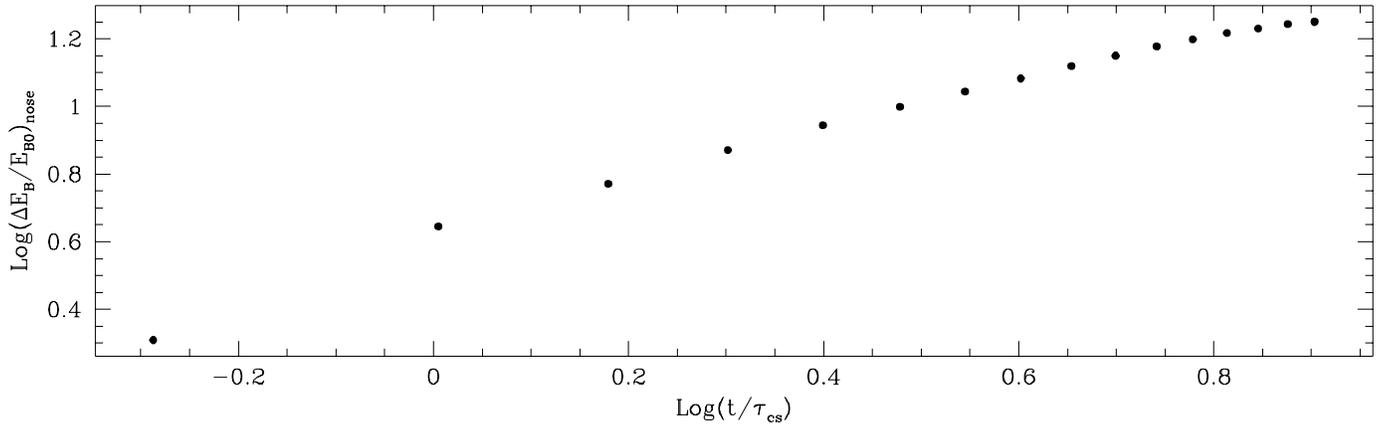
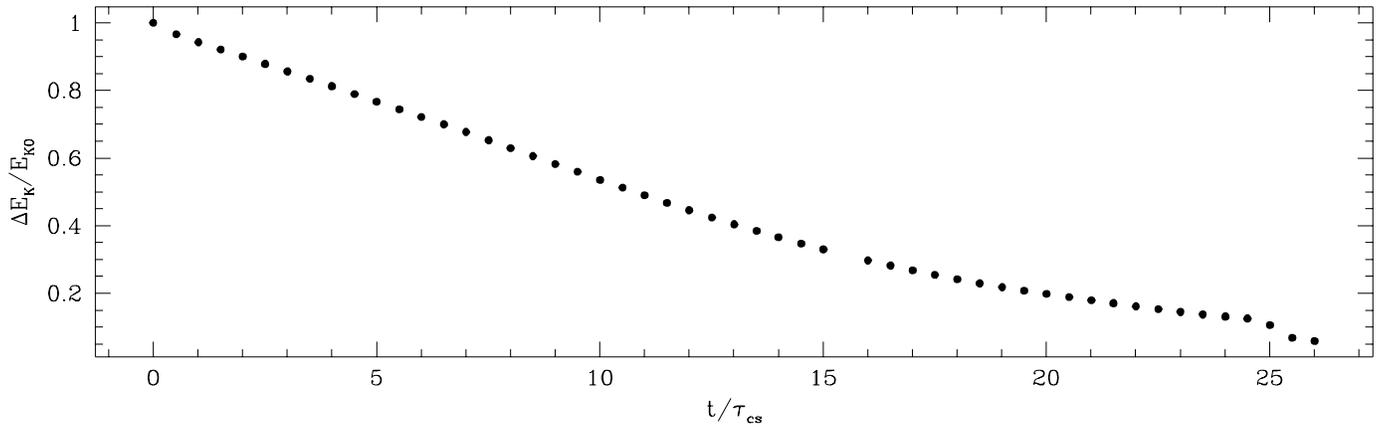
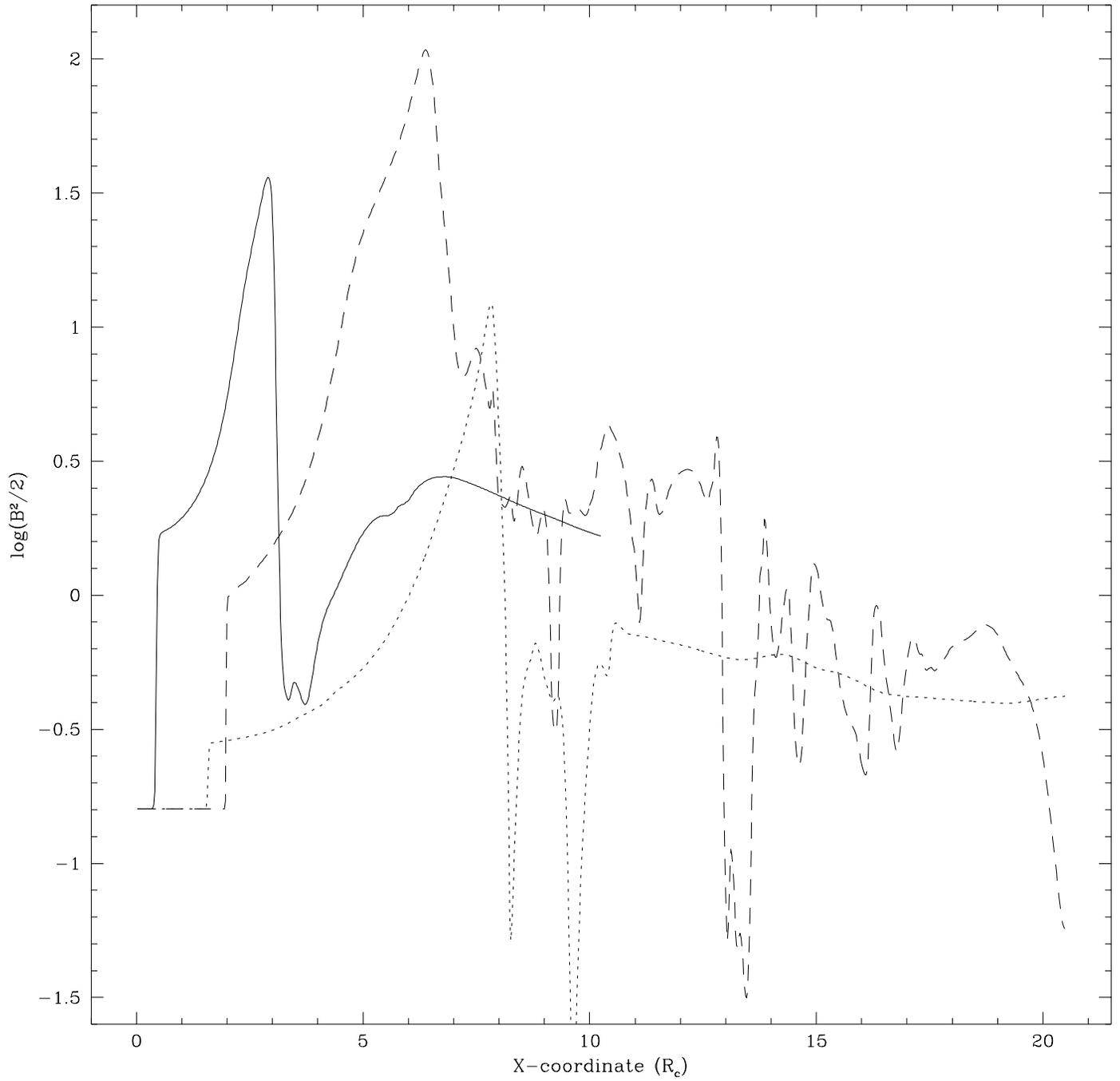
\end{document}